\documentstyle[epsf]{article}

\newcommand{\bef}{\begin{figure}}
\newcommand{\eef}{\end{figure}}
\newcommand{\be}{\begin{equation}}
\newcommand{\ee}{\end{equation}}
\newcommand{\etal}{{\it et al.}}
\newcommand{\hmp}{h^{-1}Mpc}

\def\spose#1{\hbox to 0pt{#1\hss}}
\def\ltapprox{\mathrel{\spose{\lower 3pt\hbox{$\mathchar"218$}}
 \raise 2.0pt\hbox{$\mathchar"13C$}}}
\def\gtapprox{\mathrel{\spose{\lower 3pt\hbox{$\mathchar"218$}}
 \raise 2.0pt\hbox{$\mathchar"13E$}}}
\def\inapprox{\mathrel{\spose{\lower 3pt\hbox{$\mathchar"218$}}
 \raise 2.0pt\hbox{$\mathchar"232$}}}

\begin{document}

\title{ON THE FRACTAL STRUCTURE OF THE VISIBLE UNIVERSE}


\author{L. Pietronero, M. Montuori and F. Sylos Labini\\
Dipartimento di Fisica, Universit\'a
di Roma "La Sapienza",\\
Piazzale A. Moro 2, 00185 Roma, Italy and INFM unit of Roma 1
}

\maketitle

\begin{abstract}

Some years ago we proposed a new approach for the analysis of galaxy 
and cluster correlations based on the {\it concepts and methods of 
modern Statistical Physics}. This led to the surprising result that galaxy 
correlations are fractal and not homogeneous up to the limits of the 
available catalogues. In the meantime many more redshifts have been 
measured and we have extended our methods also to the analysis of 
number counts and angular catalogues. This has led to a complete 
analysis of all the available data that we are going to present in detail 
in this lecture. In particular we will discuss the properties of the 
following catalogues: CfA, Perseus-Pisces, SSRS, IRAS, Stromlo-APM, 
LEDA, 
Las Campanas and ESP for galaxies and Abell and ACO for clusters. The 
result is that galaxy structures are highly irregular and self-similar. 
The usual
statistical methods, based on the assumption of homogeneity, 
 are therefore inconsistent for all the length scales probed until now. 
A new, more general, conceptual framework is necessary to identify
the real physical properties of these structures. In the range of 
self-similarity theories should shift from "amplitudes"
to "exponents". The new analysis shows that 
all the available data are consistent with each other and 
show fractal correlations (with dimension $D \simeq 2$) 
up to the deepest scales probed until now 
($1000 \hmp$) and even more as indicated from the new interpretation 
of the number counts. The distribution of visible matter in the 
universe is therefore fractal and not homogeneous. The evidence 
for this being very strong up to $150 \hmp$ due to the statistical robustness 
of the data and progressively weaker (statistically) at larger distances due
to the limited data.
In addition the 
luminosity distribution is  correlated with the space distribution in a 
specific way characterized by multifractal properties. 
These facts lead to fascinating conceptual implications 
about our knowledge of the universe and to a new scenario for the 
theoretical challenge in this field.
\end{abstract}

\vspace{2cm}

\section{Introduction}

This lecture reports the evidences 
for the 
fractal properties of visible matter in the universe in relation
to the debate with Prof. M. Davis. We have also included new elements
stimulated by the debate, like the analysis of 
the Stromlo-APM galaxy catalogue and a discussion 
of the angular data. 
For a comprehensive and detailed report the reader may refer 
to \cite{cp92} and to a more recent review \cite{slmp97}.

From the experimental point of view there are four main facts in 
Cosmology:
- {\it The space distribution of galaxies and clusters}: the recent availability 
of several three dimensional samples of galaxies and clusters permits 
the direct characterization of their correlation properties. - {\it The cosmic 
microwave background radiation} (CMBR), that shows an extraordinary 
isotropy and an almost perfect black body spectrum.
- {\it The linearity of the redshift-distance relation}, usually known as the 
Hubble law. This law has been established by measuring 
independently the redshift and the distance of galaxies.
- {\it The abundance of light elements in the universe}.
Each of these four points provides an independent experimental fact. 
The objective of a theory should be to provide a coherent explanation 
of all these facts together.

Our work refers mainly to the first point, {\it the space distribution of 
galaxies and clusters}  
which, however, is closely related to the 
interpretation of all the other points. In 
particular we claim that 
the usual methods of analysis are intrinsically inconsistent with
respect to the properties of these samples.
The correct statistical analysis of the 
experimental data, performed with the methods of modern Statistical 
Physics, shows that the distribution of galaxies is fractal up to the 
deepest observed scales \cite{pie87} \cite{cp92}. 
This result has caused a strong debate in the 
field because it is in contrast with the usual assumption of large scale
 homogeneity which is at the basis of most theories. Actually 
homogeneity represents much more than a working hypothesis for 
theory, it is often considered as a paradigm or principle and for some 
authors it is conceptually absurd even to question it \cite{pee93}.
For other authors instead homogeneity is just the simplest working 
hypothesis and the idea that nature might actually be more complex is 
considered as extremely interesting \cite{wei72}.
These two points of 
view are not so different after all because, if something considered 
absurd becomes real, then it is indeed very exciting. Given this 
situation it may be interesting to analyze why this question develops 
such strong feelings. This will help us to distinguish opinions from 
bare facts and to place the discussion in the appropriate perspective.
\bef
\epsfxsize 16cm
\centerline{\epsfbox{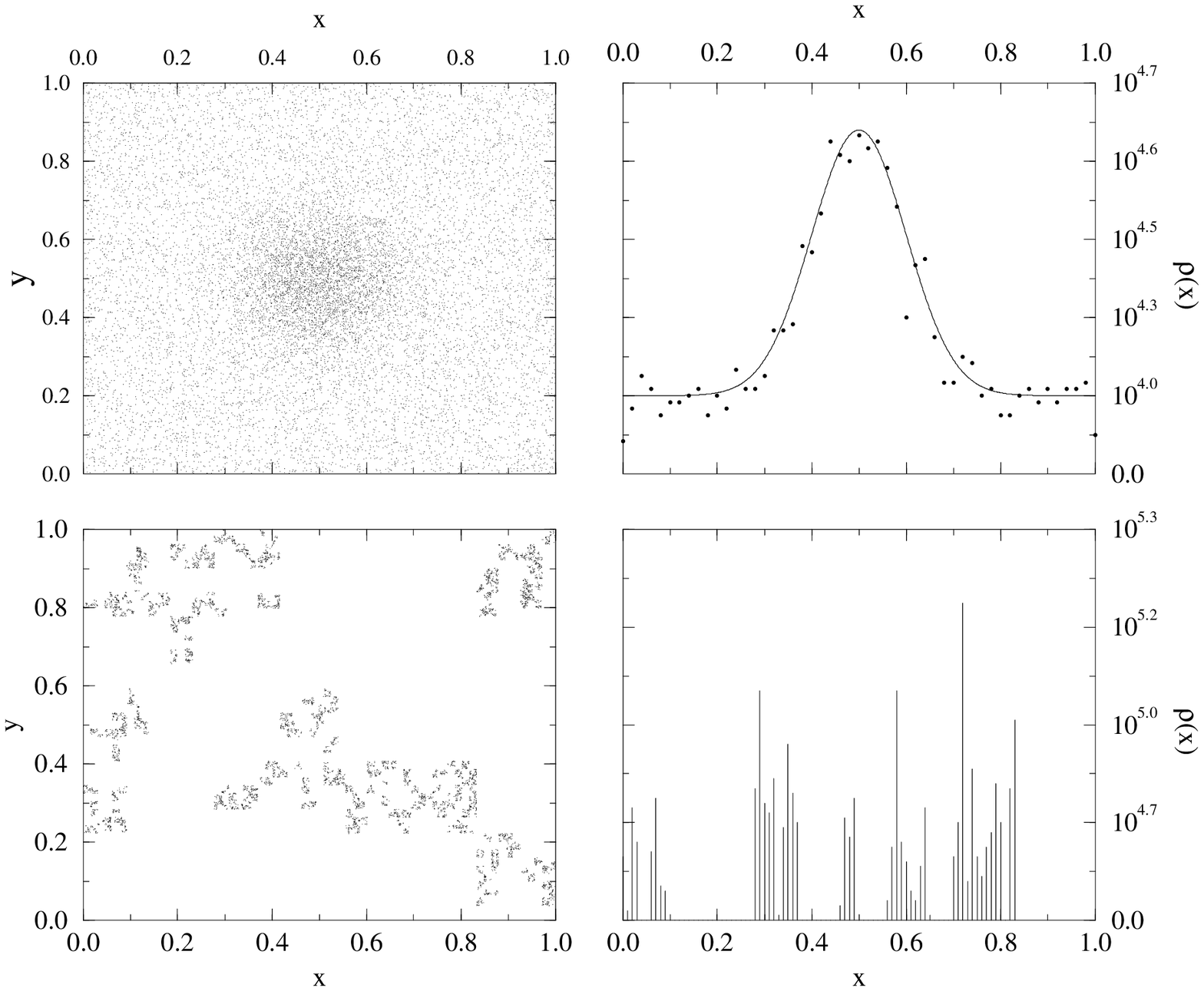}}
\caption{\label{fig1} 
Example of analytical and nonanalytic structures. {\it Top panels}
(Left)  A cluster in a homogenous distribution. (Right)
Density profile. In this case the fluctuation
 corresponds to an enhancement
of a factor 3 with respect to the average density.
{\it Bottom panels} (Left) Fractal distribution 
in the two dimensional Euclidean space. (Right) Density profile. In this 
case the fluctuations are non-analytical and there is no 
  reference value, i.e. the average density. The average density
scales as a power law from any occupied point of the structure.
}
\eef

Most of theoretical Physics is based on analytical functions and 
differential equations. This implies that structures should be 
essentially smooth and irregularities are treated as single fluctuations 
or isolated singularities. The study of critical phenomena and the 
development of the Renormalization Group (RG) theory in the 
seventies was therefore a major breakthrough \cite{wil82} \cite{amit86}.
One could observe and describe phenomena in which {\it intrinsic 
self-similar irregularities develop at all scales} and fluctuations cannot 
be described in terms of analytical functions. The theoretical methods 
to describe this situation could not be based on ordinary differential 
equations because self-similarity implies the absence of analyticity 
and the familiar mathematical Physics becomes inapplicable. In some 
sense the RG corresponds to the search of a space in which the 
problem becomes again analytical. This is the space of scale 
transformations but not the real space in which fluctuations are 
extremely irregular. For a while this peculiar situation seemed to be 
restricted to the specific critical point corresponding to the competition 
between order and disorder. In the past years instead, the 
development of Fractal Geometry \cite{man83},
has allowed us to 
realize that a large variety of structures in nature are intrinsically 
irregular and self-similar (Fig.\ref{fig1}). 

Mathematically this situation corresponds to the fact that these 
structures are singular in every point.  This property can be now 
characterized in a quantitative mathematical way by the fractal 
dimension and other suitable concepts. However, given these subtle 
properties, it is clear that making a theory for the physical origin of 
these structures is going to be a rather challenging task. This is 
actually the objective of the present activity in the field 
\cite{epv95}.
The main difference between the popular fractals like coastlines, 
mountains, trees, clouds, lightnings etc. and the self-similarity of 
critical phenomena is that criticality at phase transitions occurs only 
with an extremely accurate fine tuning of the critical parameters 
involved. In the more familiar structures observed in nature instead 
the fractal properties are self-organized, they develop spontaneously 
from the dynamical process. It is probably in view of this important 
difference that the two fields of critical phenomena and Fractal 
Geometry have proceeded somewhat independently, at least at the 
beginning.

The fact that we are traditionally accustomed to think in terms of 
analytical structures has crucial consequences on the type of questions we 
ask and on the methods we use to answer them. If one has never been 
exposed to the subtleness on nonanalytic structures, it is natural that 
analyticity is not even questioned. It is only after the above 
developments that we could realize that the property of analyticity 
can be tested experimentally and that it may or may not be present  
in a given physical system.

We can now appreciate how this discussion is directly relevant to 
Cosmology by considering the question of the Cosmological Principle 
(CP). It is quite reasonable to assume that we are not in a very special 
point of the universe and to consider this as a principle, the CP. The 
usual mathematical implication of this principle is that the universe 
must be homogeneous. This reasoning implies the hidden assumption 
of analyticity that often is not even mentioned. In fact the above 
reasonable requirement only leads to {\it local isotropy}.
 For an analytical 
structure this also implies homogeneity \cite{wei72}. However, if 
the structure is 
not analytical, the above reasoning does not hold. For example a 
fractal structure has local isotropy but not homogeneity. 
In simple terms one observes the same mix of structures and voids 
in different directions (statistical isotropy).
This means 
that a fractal structure satisfies the CP in the sense that all the points 
are essentially equivalent (no center or special points) but this does 
not imply that these points are distributed uniformly. In this respect 
the saturation of the dipole moment cannot be considered as a test of 
homogeneity because this property is also present in fractal structures 
\cite{sl94}.

This important distinction between isotropy and homogeneity 
has not been adequately appreciated in M. Davis' lecture, and it has 
other important consequences. For example it clarifies that drawing 
conclusions about real galaxy correlations from the angular 
distributions alone can be rather misleading as we are going to see in detail later. 
In addition, from this 
new perspective, the isotropy of the CMBR may appear less 
problematic in relation to the highly irregular three dimensional 
distribution of matter and this may lead to theoretical approaches of 
novel type for this problem.
In the present work, however, we are going to limit our discussion to 
the optimal way to analyze the data provided by the galaxy catalogues 
from the broader perspective in which analyticity and homogeneity 
are not assumed a priori but they are explicitly tested. The main 
result is that the data become actually consistent with each other and 
coherently point to the same conclusion of fractal correlations up to 
the present observational limits. In this lecture we present a colloquial 
discussion of the main results and their implications.

\section{ Statistical Methods and Correlation Properties}

{\it - Usual arguments:} 
Before the extensive redshift measurements of the 80s the 
information about the galaxy distribution was only in terms of the two 
angular coordinates. These angular distributions appeared rather 
smooth at relatively large angular scale, like for example the lower 
part of Fig.\ref{fig2}.
\bef
\epsfxsize 12cm
\centerline{\epsfbox{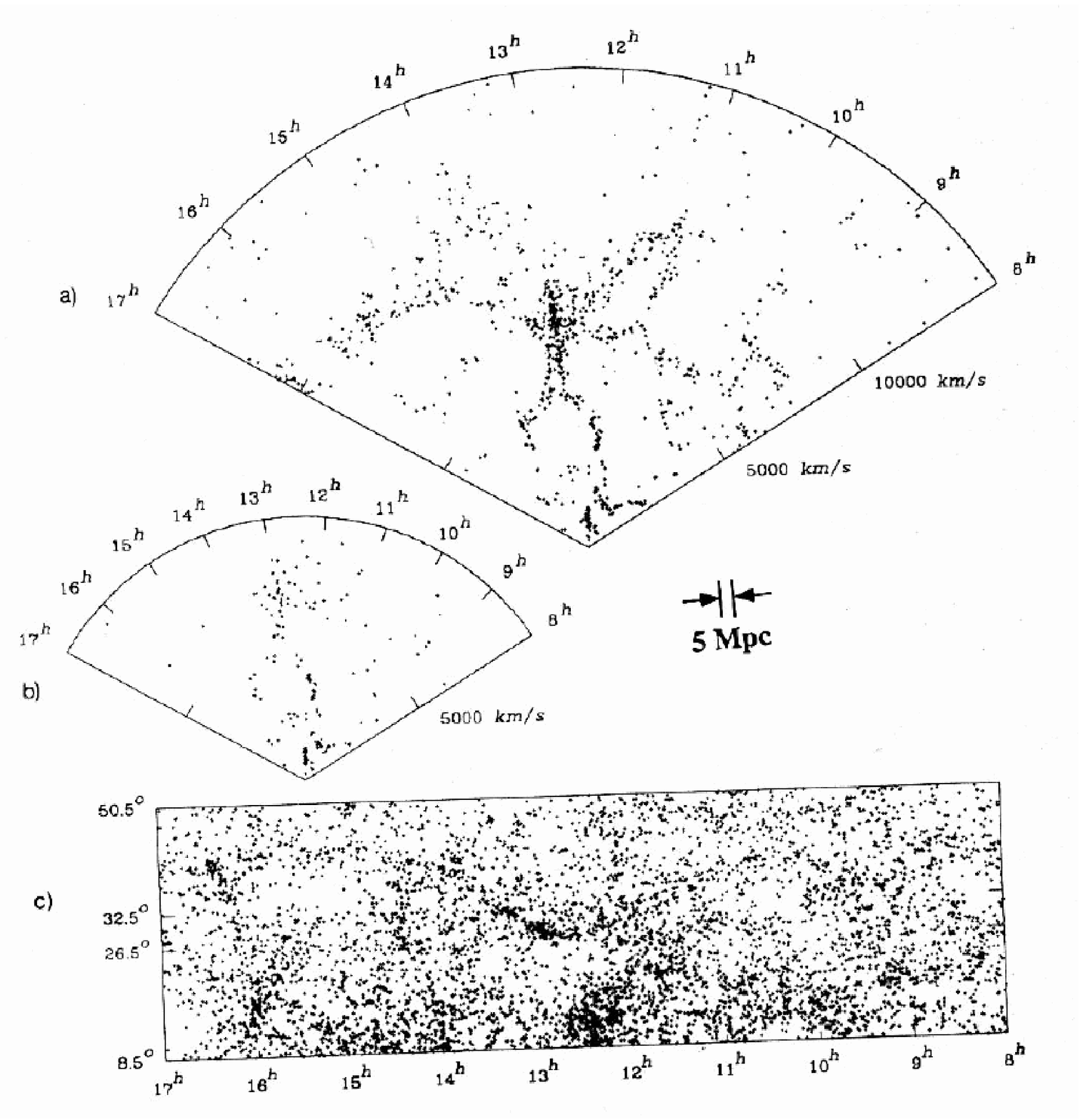}}
\caption{\label{fig2}
A slice of  the three dimensional galaxy distribution (a) (b)
compared with the corresponding (c)
angular distribution (the portion between 
$26.5^{\circ}$ and $32.5^{\circ}$ - [11]).
Note that the angular distribution appears relatively 
homogenous while the real three distribution in space is much more irregular. 
In particular this picture shows the so-called Great Wall 
which extends over the entire sample (at least $170 \hmp$).
Note that it is impossible to 
identify structures of this typo from the angular data alone.
We also show the size of the galaxy "correlation-length" ($r_0 = 5 \hmp$)
derived from the standard analysis. The more 
general analysis we discuss  here implies instead that 
an eventual correlation length should be larger than any observable
structure, i.e. $\gg 170 \hmp$ and that the present data show well defined
fractal properties up to the sample limits. 
These pictures clarify the intrinsic ambiguity of the angular catalogs in 
relation to the arguments of Prof. Davis. 
They show that a fractal 
distribution in 3-d (top) projects in a smooth angular distribution 
(bottom). No one could have predicted the real 3-d properties 
from the angular distribution alone. }
\eef
Assuming that this smoothness corresponds to a real homogenization 
in 3-d space and estimating the characteristic depth of the angular 
catalogue from the magnitudes  "a characteristic length"   
$r_0 = 5 \hmp$ has been estimated \cite{pee93}. 
The idea was that beyond such a distance  the 3-d galaxy 
distribution would become gradually homogeneous and it could be well 
approximated by a constant galaxy density. This average density, apart from the 
eventual Dark Matter, would be the one to use into Einstein equations 
to derive the Friedmann metric and the other usual concepts. 

Later, the 
measurements of the galaxy redshifts plus the Hubble law provided 
also the absolute distances and could identify the position of galaxies 
in space. However the 3-d galaxy distributions turned out to be much 
more irregular with respect to their angular projections and showed 
{\it large structures and large voids},
 as shown in the upper part of Fig.\ref{fig2}. 
At first these irregular structures appeared to be in contradiction 
with the picture derived from the angular catalogues and, as we are 
going 
to see, they really are. However in 1983 
 a correlation analysis of the 3-d distribution 
(CfA1 catalog \cite{huc83}) was performed 
by Davis and Peebles \cite{dp83} 
and the 
result was again that the correlation length was $r_0 = 5 \hmp$ as for 
the angular catalogues. This seemed to resolve the puzzle because it 
was interpreted as if a relatively small correlation length can be 
consistent with the observation of large structures. This value for
 $r_0$ 
has not been seriously questioned even after the observation of huge 
structures, like the galaxy wall, that extend up to $\sim 170 \hmp$. 
Fig.\ref{fig2} is a clear example 
of the smoothing effect of angular projections, and 
it already gives a clear indication of the 
compatibility of a fractal structure in 3-d
with a smoother projection.

The usual correlation analysis is performed by estimating at which 
distance ($r_0$) 
the density fluctuations are comparable to the average density in the 
sample ($\xi(r)=  \langle n(0)n(r) \rangle/ \langle n \rangle^{2} -1$; 
$\xi(r_0) \equiv 1$). 
Now everybody agrees that there are fractal correlations at least at 
small scales. The important physical question is 
therefore to identify the 
distance $\lambda_0$ 
at which, possibly, the fractal distribution has a 
crossover into a homogeneous one. This would be the
 {\it real correlation length }
beyond which the distribution can be approximated by an average 
density. The problem is therefore to understand the relation between $r_0$
 and $\lambda_0$: are they the same or closely related or do 
they correspond to different properties? 
This is actually a subtle point with respect to the concepts discussed in 
the introduction. In fact, if the galaxy distribution becomes really 
homogeneous at a scale $\lambda_0$   within the sample in question, 
then the value of $r_0$
 is related to the real correlation properties of the distribution and one 
has $r_0 \simeq \lambda_0/2$. 
If, on the other hand, the fractal correlations extend up to the sample 
limits, then the resulting value of $r_0$ has nothing to do with the real 
properties of the galaxy distribution but it is fixed just by the size of the 
sample \cite{cp92}. 

{\it New Perspective:} Given this situation of ambiguity with respect to the real 
meaning of $r_0$ it is clear that the usual correlation study in terms of 
the function $\xi (r)$ is not the appropriate method to clarify these 
basic questions.
The essential problem is that, by using the function $\xi(r)$, one 
defines the amplitude of the density fluctuations by normalizing them 
to the average density of the sample in question. This implies that the 
observed density should be the real one and it should not depend on 
the given sample or on its size apart from Poisson fluctuations. 
However, if the distribution shows long range (fractal) correlations, this 
approach becomes meaningless. For example if one studies a fractal 
distribution with $\xi(r)$ a characteristic length $r_0$
 will be identified, but this is clearly an artifact because the structure is 
characterized exactly by the absence of any defined  length \cite{cp92}.

The appropriate analysis of pair correlations should therefore be 
performed using methods that can check homogeneity or fractal 
properties without assuming a priori either one.
The simplest method to do this is to consider directly  the conditional 
density  $\Gamma(r) \sim  \langle n(0)n(r) \rangle$ without comparing it 
to the average. There are several other methods that have been 
discussed elsewhere \cite{cp92} \cite{slmp97}. 
This is not all however because one has also to be 
careful not to make hidden assumptions of homogeneity in the specific 
procedure to evaluate these correlations. For example, if a galaxy is 
close to the boundary of the sample, it is possible that the sphere of 
radius $r$ around it, {where the conditional density is computed}, 
 may lie in part outside the sample boundary. In this 
case the usual procedure is to use weighting schemes of various types 
to include also these points in the statistics. In this way one implicitly 
assumes that the fraction of sphere contained in the sample is 
sufficient to estimate the properties of the full sphere. This implies 
that the properties of a small volume are assumed to be
the same as for a larger 
volume (the full sphere). This is a hidden assumption of homogeneity 
that should be   avoided by including only the properties of those 
points for which a surrounding sphere of radius $ r$ is fully included in 
the sample. These procedures are fully standard in modern statistical 
mechanics and a detailed description can be found in \cite{cp92}. 
 This means that the statistical validity of a sample is limited to the 
radius of the largest sphere that can be contained in the sample. We 
call this distance $R_s$ and it should not be confused with the sample 
depth $R_d$, which can be in general much larger, depending on the survey geometry.

In 1988 we  reanalyzed the CfA1 catalogue \cite{cps88}.
 The 
result was that the catalogue has   statistical validity up to 
$R_s = 20 \hmp$  and, up this length, it shows well defined fractal 
correlations. This shows 
therefore that the "correlation length" $r_0=5\hmp$
 derived by \cite{dp83} 
was a spurious result due to an 
inappropriate method of analysis and it has nothing to do with the real 
correlation properties of the system. A similar analysis of the Abell 
cluster catalogue also showed fractal properties up to $R_s = 80\hmp$
 so  that also the cluster "correlation length" 
$r_0^c = 25\hmp$ \cite{bs83}
 should be considered as spurious. One consequence of 
these results was that the so called galaxy-cluster mismatch could be 
automatically eliminated by the appropriate analysis. Also other properties 
like $\delta N/N$, directly related to $r_0$, 
suffer from the same consistency problems because of the lack of a 
reference value \cite{bslmp94}. This situation 
led to a rather controversial debate in the field. In the meantime 
many more data have became available and we have performed a 
complete analysis of all the data for galaxies and clusters. In the 
following we report the main results.

\section{Analysis of the Galaxy Distributions}

Here we discuss the correlation properties of the galaxy distributions 
in terms of volume limited catalogues \cite{cp92}
 arising from most of  the  
50.000 redshift measurements that have been made to date. A first 
important result will be that {\it the samples 
are statistically rather good }
and their properties are in agreement with each other. This gives a 
new perspective because, using the standard methods of analysis, the 
properties of different samples appear contradictory with each other 
and often this is considered to be a problem of the data (unfair 
samples) while, we show that this    is due to the inappropriate methods 
of analysis.
In addition essentially all the catalogues show well defined fractal 
correlations up to their limits and the fractal dimension is $D \simeq 2$. The
 few exceptions to this result will be discussed and interpreted in 
detail.

\begin{table}
\caption{\label{tab1}The volume limited catalogues are characterized by the following 
parameters:
- $R_d (\hmp)$ is the depth of the catalogue
- $\Omega$ is the solid angle
- $R_s (\hmp)$ is the radius of the largest sphere 
that can be contained in the catalogue volume. 
This gives the limit of statistical validity of the sample.
- $r_0(\hmp)$  is the length at which $\xi(r) \equiv 1$.
- $\lambda_0$ is the eventual real crossover to a homogeneous 
distribution that is actually never observed. 
The value of $r_0$ 
is the one obtained 
in the deepest sample. 
The CfA2 and SSRS2 data are not yet available.
See Ref [13], [17]-[23] and [31] 
(distance are expressed in $\hmp$).
}
\begin{tabular}{|c|c|c|c|c|c|c|}
\hline
     &      &          &    &              &  &       \\
\rm{Sample} & $\Omega$ ($sr$) & $R_d  $ & $R_s  $ 
& $r_0  $& $D$ & $\lambda_0 $ \\
     &       &    &    &    &  		   &       \\
\hline
CfA1 & 1.83  & 80 & 20 & 6  & $1.7 \pm 0.2$& $>80$    \\
CfA2 & 1.23  & 130& 30 & 10 & 2.0          & $ ? $    \\
PP   & 0.9   & 130& 30 & 10 & $2.0 \pm 0.1$& $> 130$  \\
SSRS1& 1.75  & 120& 35 & 12 & $2.0 \pm 0.1$ & $ >120$      \\
SSRS2& 1.13  & 150& 50 & 15 & 2.0          & $?$      \\
Stromlo-APM& 1.3  & 100& 30 & 10 & $2.2 \pm 0.1$& $ > 150$      \\
LEDA & $4 \pi$ & 300& 150& 45 & $2.1 \pm 0.2$& $>150$   \\
LCRS & 0.12  & 500& 18 & 6  & $1.8 \pm 0.2$& $> 500$    \\
IRAS $1.2 Jy$ & $4 \pi$ & 80 & 40 & 4.5& $2.0 \pm 0.1$& $\simeq 25$\\
ESP  & 0.006 & 700& 10 & 5  & $1.9 \pm 0.2$& $>800$   \\
     &       &         &       &      &  &       \\
\hline
\end{tabular}
\end{table}

\bef
\epsfxsize 10cm
\centerline{\epsfbox{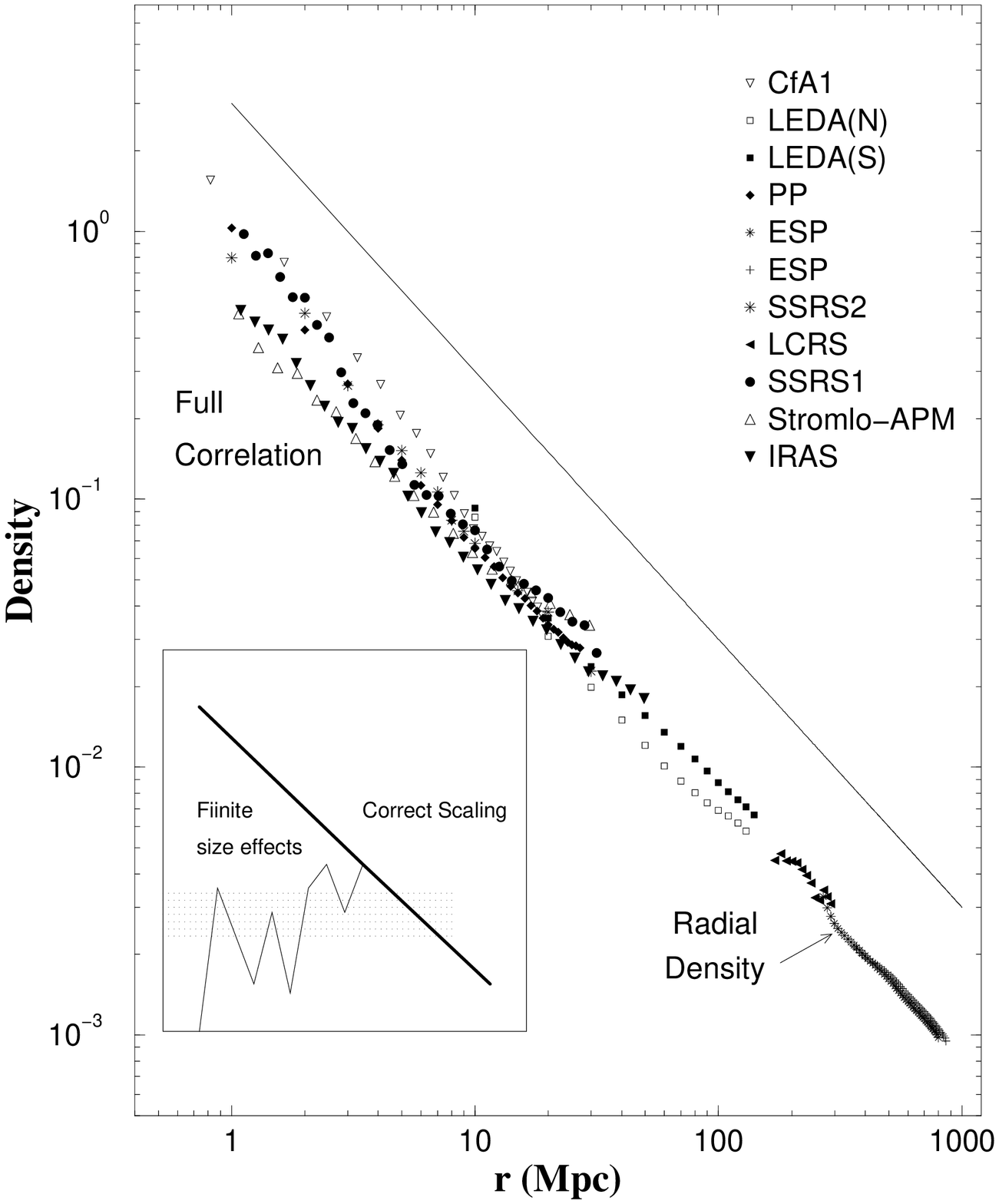}}
\caption{
\label{fig3}
Full correlation for the various available
redshift  catalogues in the range of distances
$0.1 \div 1000  \hmp$. A reference line with a slope 
$-1$ is also shown (i.e. fractal dimension $D = 2$).
Up to $\sim 150 \hmp$ the density is computed by the full correlation 
analysis, while above $\sim 150 \hmp$ it is computed through the 
radial density.
For the full correlation the data of the various catalogues are normalized
with the luminosity function and they match very well 
with each other [2]. This is an important test of the statistical validity and 
consistency of the various data.
In the {\it insert panel} it is shown the 
schematic behavior of the radial 
density versus distance computed from the vertex (see text) [27].
The behaviour of the radial density allows us to extend the power law
correlation up to $\sim 1000 \hmp$. However a rescaling 
is necessary to match the radial density to the 
conditional density [2].
}
\eef
\bef
\epsfxsize 10cm
\centerline{\epsfbox{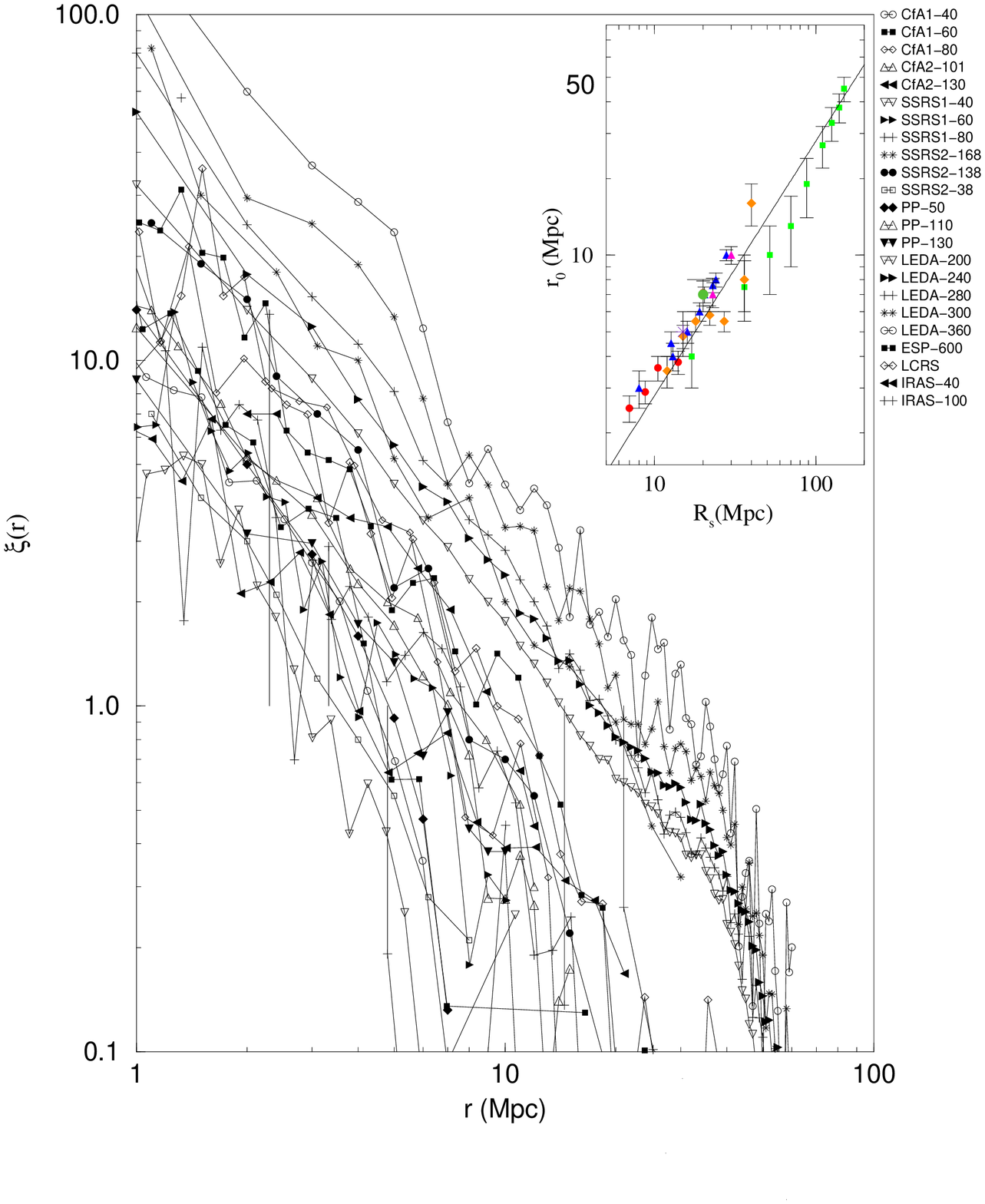}}
\caption{\label{fig4} 
Usual analysis based on the 
function $\xi(r)$ of the same galaxy catalogues of Fig.3. This 
analysis is based on the 
a priori and untested assumption 
of the analyticity and homogeneity.
These properties are not present in the 
real galaxy distributions and the results appear 
therefore rather confusing. This lead to the 
impression that galaxy catalogues are not 
good enough and to a variety of theoretical 
problems like the galaxy cluster 
mismatch, luminosity 
segregation, the linear and non linear evolution, etc.. 
The situation changes completely and it becomes rather 
clear if one adopts 
the more general framework that is at the basis of Fig.3. 
In the {\it insert panel} we show the dependence 
of $r_0$ on $R_s$ for all the catalogs. The linear behaviour is a consequence 
of the fractal nature of galaxy distribution in these samples [2].
}
\eef 
The main data of our correlation analysis 
are collected in Fig.\ref{fig3} (left part)
 in which we report the 
{\it conditional density as a function of scale}
 for the various catalogues. 
The relative position of the various lines is not arbitrary but it is fixed 
by the luminosity function, a part for the cases of 
IRAS and SSRS1 for which this is 
not possible. The properties derived from different 
catalogues are compatible with each other and show a {\it power law 
decay} for the conditional density from $1 \hmp$ to $150 \hmp$
 without any tendency towards homogenization (flattening). This 
implies necessarily that the value of $r_0$ (derived from the $\xi(r)$ 
approach) will scale with the sample size $R_s$ as shown also from the 
specific data about $r_0$ of the various catalogues 
\cite{cps88} \cite{pp96}. The 
behaviour 
observed  corresponds to a fractal structure with dimension $D \simeq 
2$. The smaller value of CfA1 was due to its limited size. An 
homogeneous distribution would correspond to a flattening of the 
conditional density which is never observed.  Usually the $\xi(r)$ approach 
also leads to a smaller value of $D$ (or a larger value of $\gamma=3-
D$) as derived from the small scale properties. This is due to the fact 
that the fit is made close to $r_0$ and it is affected by the fact that 
$\xi(r)$ (in log coordinates) is becoming steeper because it is 
approaching the value zero.

It is interesting to compare the analysis of Fig.\ref{fig3} 
with the usual one, made with the function $\xi(r)$, 
for the same galaxy catalogues. This is 
reported in Fig.\ref{fig4} and, from this point of view,
the various data appear to be in strong disagreement with 
the each other. This is due to the fact that the 
usual analysis looks at the data from the 
perspective of  analyticity and large scale homogeneity 
(within each sample). These properties are never tested and they 
are actually 
not present in the real galaxy distributions so the result is 
rather confusing (Fig.\ref{fig4}).
Once the same data are analyzed within a broader 
perspective the situation becomes clear (Fig.\ref{fig3}) and the data 
of different catalogues result in agreement with each other.
In addition in the insert of Fig.\ref{fig4} we show the dependence 
of $r_0$ on $R_s$ for all the catalogs. The linear behaviour is a consequence 
of the correlation properties of Fig.\ref{fig3} and it provides 
an additional evidence of fractal behaviour to all scales.
In this respect, the proposed luminosity bias effect mentioned by M. Davis 
appears essentially irrelevant while, on the contrary, the linear dependence 
of $r_0$ on $R_s$ is very clear.

The new picture allows us to make clear 
predictions for the value of $r_0$ for the forthcoming catalogs CfA2 and SLOAN. 
Considering the predicted behaviour of $r_0 \simeq 0.3 R_s$ (for $D \simeq 2$ 
- see insert of Fig.\ref{fig4}) we expect that in CfA2 one should have 
$r_0 \approx 15 \div 20   \hmp$ while for the full SLOAN catalog 
$r_0 \approx  50 \div 60 \hmp$ (of course these values depend on the solid 
angle of the survey and the volume limited sample considered 
in the way precisely discussed previously).
We stress however that these predictions refer to the full solid angle catalogs
and for subsamples one should consider {\it the corresponding value of
$R_s$. }

It is important to remark that analyses like those of 
Fig.\ref{fig3} have had a profound influence in the field in various 
ways: first the various catalogues appear in conflict with each 
other. This has generated the concept of "{\it not fair sample}" 
and a strong mutual criticism about the validity of the 
data between different groups. In other cases the 
discrepancy observed in Fig.\ref{fig4} have been considered as 
real physical problems for which various theoretical 
approaches have been proposed. These problems are, 
for example, the galaxy-cluster mismatch, luminosity segregation, 
the richness clustering relation and the linear 
and non linear evolution of the perturbations 
corresponding to the "{\it small}" or 
"{\it large}" amplitudes of fluctuations.
We can now see that all this problematic is not real and 
it arises only from a statistical analysis based on inappropriate and to 
restrictive assumptions that do not find a 
correspondence in the physical reality. It is also important 
to note that, even if the galaxy distribution 
would eventually become homogeneous at some large scale, the use of the 
above statistical concepts is anyhow inappropriate for the 
range of scales in which the 
system shows fractal correlations as those shown in Fig.\ref{fig3}.

Contrary to the claims of  Prof. Davis we would like to stress  that a
fractal distribution has a very strong property: {\it it shows
global power-law
correlations up to the sample depth}. Such correlations {\it cannot be due}
neither 
to  an inhomogeneous sampling of an homogeneous distribution, 
nor to some selection effects that may occur in the observations. 
Namely, suppose that  a certain kind of sampling reduces the 
number of galaxies as a function of distance.
 Such an effect, in no way can lead to long range correlations,
because when one computes $\Gamma(r)$, one makes an average 
over all the points inside the survey.
In any case this possible bias could be detected 
by a difference in the values of $\Gamma(r)$ at different depths, which is 
not observed \cite{slmp97}.
We observe instead that all the catalogues, independently  on 
their completeness, show precisely the same correlation properties.

In connection with the conditional density decay of Fig.\ref{fig3},
it is interesting to note that the Hubble law (redshift versus distance)
has been experimentally tested in the same range of scales. 
These two experimental facts show that the Hubble law is compatible 
with a fractal universe, contrary to the usual theoretical interpretation 
\cite{bslmp94} \cite{bpslt96}.

In Fig.\ref{fig5} 
\bef
\epsfxsize 16cm
\epsfysize 15cm
\centerline{\epsfbox{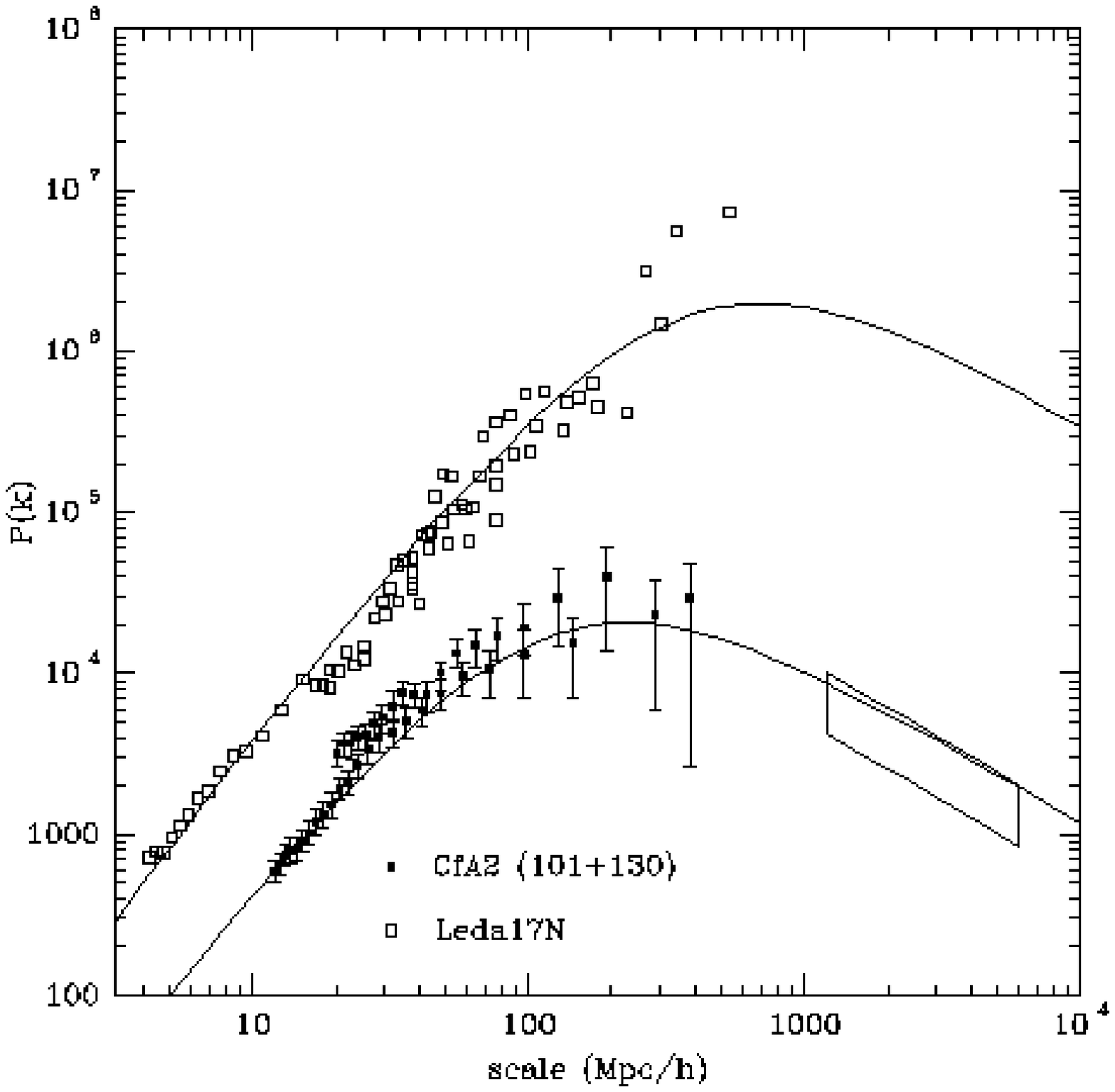}}
\caption{\label{fig5}
 Power spectrum (PS) of several samples of LEDA
together with the CfA2 power spectra (from [22]
The amplitude of the PS
depends on the sample depth, and it is 
larger in the case of LEDA, because in this case 
the effective depth is larger than the CfA2 one 
(see Table 1).
The curved shape of the PS is also 
 spurious  that does not
correspond to any intrinsic feature
in the case of fractals, but it is only due to the 
finiteness of the sample volume [25].
We also plot a CDM-like fit both to 
the CFA2 data and the LEDA data.
On the right, we plot the COBE data. 
No reasonable CDM curve can fit the data and COBE 
(from [25])}
\eef
we report the power spectrum analysis (Fourier conjugate of $\xi(r)$) 
of some catalogues. 
Also in this case, as for $\xi(r)$,
there is a specific dependence of the 
properties on the sample size, in full agreement with the direct 
correlation analysis of Fig.\ref{fig3}.
In particular, for a fractal structure, 
 the amplitude of the power spectrum 
is a function of the sample size and the shape 
is characterized by a 
turnover: both these features, bending  and scaling,
are a manifestation of 
the finiteness of the survey volume, and 
cannot be
interpreted as the convergence to homogeneity, nor to a power spectrum
flattening.
 A detailed discussion of the power 
spectrum can be found in \cite{sla96}.

Essentially similar results have been obtained for the Abell and ACO 
cluster catalogues \cite{cp92} \cite{msla97} \cite{slmp97}. 
Also in these cases we observe a power law 
behaviour for the cluster 
correlations with $D \simeq 2$ and without any 
tendency towards homogenization up to $\sim 80 \hmp$.

All these results imply that the previous "correlation lengths" of $5 \hmp$
and 
$25 \hmp$, introduced for galaxies and clusters, are spurious and no 
real correlation length can be defined from the data. 
Therefore the much discussed mismatch between galaxy and cluster 
correlations, that is also at the basis of various theories for structure 
formation, does not actually exist. Cluster correlations correspond just 
to the continuation of galaxy correlations at larger scales. In the 
language of 
Statistical Physics, cluster catalogues are the coarse grained 
version of galaxy catalogues.

\section{Statistical Validity of Catalogues}

Often the concept of {\it fair sample} has been used to mean 
homogeneous sample. For this reason various samples are declared as 
{\it not fair} just because they are not homogeneous. We have seen 
instead that it was the method of analysis to be inappropriate while 
the samples are actually rather good in a statistical sense. This means 
that one can derive from them a statistically valid information about 
their correlation properties. 

In relation to the statistical validity it is interesting to consider the 
IRAS catalogues  because they seem to differ from all the other ones and 
to show some tendency towards homogenization at a relatively small scale 
Actually the point of apparent homogeneity 
is only present in some samples, it varies 
from sample to sample between $\sim 15 \div 25 \hmp$ 
and it is strongly dependent on the dilution of the sample.
Considering that structures and voids are much 
larger than this scale and that the IRAS galaxies appear to be just 
where luminous galaxies are it is clear that this tendency appears 
suspicious. One of the characteristic of the IRAS catalogues with respect 
to all the other ones is an extreme degree of dilution: this catalogue 
contains only a very small fraction of all the galaxies. It is important  
therefore to study what happens to the properties of a given sample if 
one dilutes randomly the galaxy distribution up to the IRAS limits. A 
good test can be done by considering the Perseus Pisces catalogue and 
eliminating galaxies from it. The 
original distribution shows a well defined fractal behavior. 
By diluting it to the level of IRAS one observes an artificial flattening 
of the correlations \cite{slgmp96} (see Fig.\ref{fig6}).
\bef
\epsfxsize 14cm
\centerline{\epsfbox{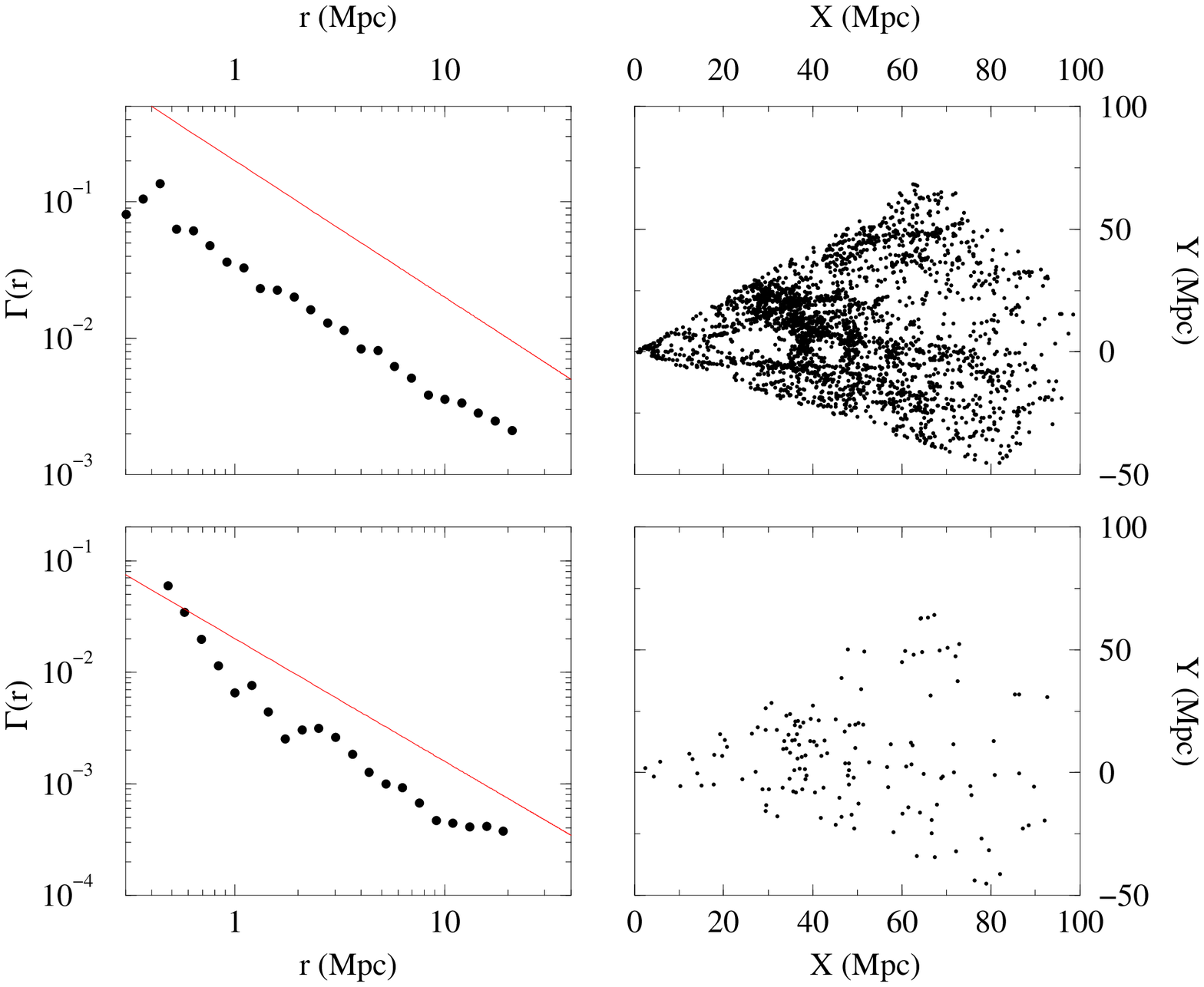}}
\caption{ \label{fig6}
{\it Top panels:} The conditional density (left) for a volume limited sample
of the full 
Perseus-Pisces redshift survey (right). 
The percentage of galaxies present in the sample is $\sim 6 \%$.
The slope is $-\gamma=-1$.
{\it Bottom panels:} In this case the percentage of galaxies is $\sim 1 \%$,
and the number of galaxies is the same of the IRAS $1.2 Jy$ sample 
in the same region of the sky. We can see that at small scale 
we have a $1/r^3$ decay just due to the sparseness of the sample,
while at large scale the shot noise of the sparse sampling 
overcomes the real correlations
and produces an apparent trend to homogenization [27].
In our opinion, this effect, due to sparsness of 
this sample, is the origin of the apparent trend 
towards homogeneization observed in some of the 
IRAS samples.}
\eef
 This effect does not correspond to a real 
homogenization but it is due to the of dilution. In fact it can be 
shown that when the dilution is such that the average distance 
between galaxies becomes comparable with the largest voids 
(lacunarity) of the original structure there is a loss of correlation 
and the shot noise of the sparse sampling 
overcomes the real correlations
and produces an apparent trend to homogenization [27].
This allows us to reconcile this peculiarity of the 
IRAS data with the properties of all the other catalogues.
Analogous considerations for other sparse samples 
like QDOT and the Stromlo-APM samples \cite{slmp97}.

The correlations discussed up to now are well defined 
statistically but limited to the radius 
of the largest sphere that can be 
contained in the sample $R_s$. 
For example for Las Campanas the depth is 
very large $R_d \sim 500\hmp$
 but $R_s$ is only $20 \hmp$ because the sample is very thin. 
So, it is not surprising that the value of $r_0$ is also small 
($6 \div 7 \hmp$). Given this 
situation it would be very interesting to find some 
method that is limited by $R_d$ instead of $R_s$.

Galaxy samples have typically a conic shape and, in this respect, they 
are three dimensional objects. Considering a volume limited sample, if 
one simply counts the number of galaxies within a distance $r$ from 
the vertex this number should go like $r^3$ for a homogeneous 
distribution and like $r^D$ for a fractal one. In order to compare with 
the previous correlation analysis it is actually convenient to consider 
the density instead of the total number. The problem with such an 
analysis is that one cannot average from different observational points 
but the count is from a single point, the vertex. This situation 
corresponds to a reduced statistical quality that should be carefully 
analyzed. 

At very small distances we are not going to find any point. 
When a distance of the order of the minimal one between galaxies is 
reached, we begin to have a signal but this is strongly affected by 
finite size effects as shown schematically in the insert of Fig.\ref{fig3}. 
Eventually at some distance $\lambda$, the number of points becomes 
large enough that one reaches the correct scaling behavior. It can be 
shown that this characteristic length for the statistical significance of 
this method is about ten times the minimal distance between galaxies. 
For various catalogues the value of $\lambda $ is appreciably smaller that 
$R_d$ and, in these cases, a useful information can be obtained for the 
length scales between $\lambda$ and $R_d$. A detailed discussion of this 
method can be found in \cite{slgmp96}.
This approach is 
quite useful  because it allows us to use the thin deep 
catalogues up to their total depth $R_d$. Particularly interesting in this 
respect is the ESP catalogue whose depth 
extends up to $900\hmp$. Also
 for the Las Campanas sample it is possible to obtain some 
  information despite the peculiar and unfortunate 
luminosity selections of this catalogue. The results of these deep 
catalogue are reported in Fig.\ref{fig3}
 together with those of the other 
catalogues discussed before. 
The amplitude of the density computed from the vertex 
for all the catalogues is systematically shifted by about a factor of 3 
(that has been rescaled in Fig.\ref{fig3})
with respect to the full correlation analysis. This shift is due 
to the fact that usually, observations 
point towards zones that are rich of 
galaxies. A detailed discussion of this effect is reported in \cite{slmp97}.

The behaviour of the density decay from the vertex shows the same 
power law behavior of the full correlation analysis of Fig.\ref{fig3}
(left part)  but this 
property is now shown to extend up to $900\hmp$. The agreement 
between different catalogues and different methods of analysis is 
remarkable and it shows a well defined fractal behavior for the galaxy 
correlations extending from $1$ to about $1000\hmp$ (Fig.\ref{fig3}).
This analysis refutes therefore the early comments about a 
possible homogenization in Las Campanas, based on the visual impression of the 
data. 

In relation to the lecture by Prof. Davis, it should be noted that the function
$n(z)$ (redshift counts in a magnitude limited sample) it 
can be shown \cite{slmp97} that the behaviour of such a quantity
is mostly related to the luminosity selection function of the survey
rather than to the behaviour of the space density. In particular,
 such a quantity
cannot be used to distinguish between 
fractal or homogeneous 
 properties of the 
galaxy distribution.

\section{Number Counts and Angular Correlations}

The above discussion of the density decay from the vertex brings us 
naturally to the 
problem of the galaxy number counts that is also 
performed in this way, namely by counting from the origin. 
There are however some relevant differences 
because all the properties discussed up to here refer to volume limited 
samples while the galaxy counts are defined by their apparent 
magnitude. The behaviour of the number versus magnitude relation 
($N(<m)$) is reported in Fig.\ref{fig7}. 
\bef
\epsfxsize 10cm
\centerline{\epsfbox{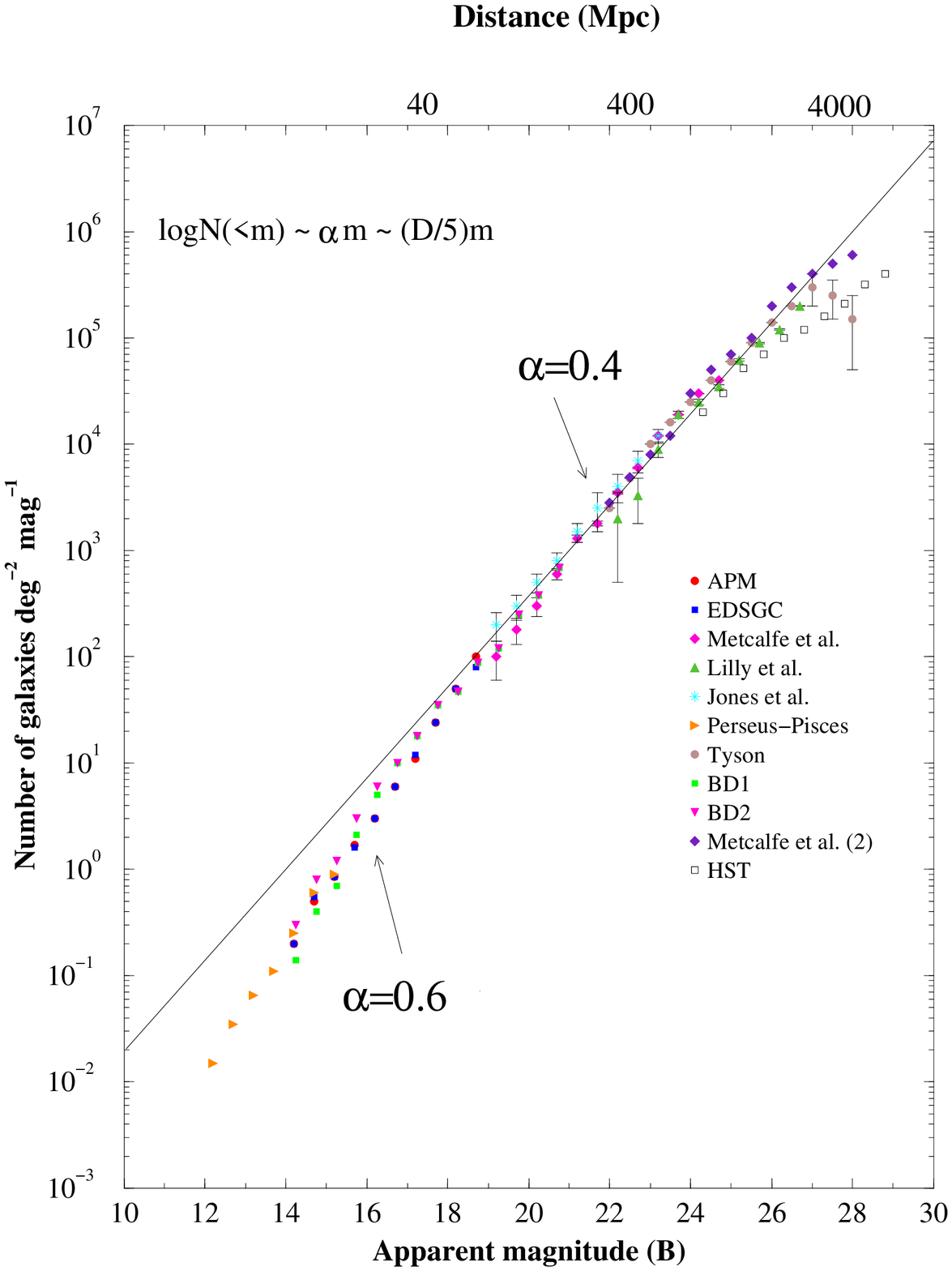}}
\caption{ \label{fig7}
The galaxy number counts in the $B$-band,
from several surveys.
In the range $12 \ltapprox  m \ltapprox 19$
the counts show an exponent 
$\alpha \simeq 0.6 \pm 0.1$, while
in the range $19 \ltapprox m \ltapprox 28$ 
the exponent is $\alpha  \simeq 0.4$.
The amplitude of the galaxy number counts for $m \gtapprox 19$
(solid line)  is computed from the determination of the 
prefactor $B$ of the density $n(r) = Br^{D-3}$ (with $D=2$ - see text) 
at small scale
and from the knowledge of the galaxy luminosity function. The distance 
is computed for a galaxy with $M=-16$ and we have used 
$H_o=75 km sec^{-1} Mpc^{-1}$ [27] [2].}
\eef
The exponent $\alpha$ defined by this relation can be related to the 
dimension $D$ of the galaxy distribution ($\alpha = D/5$).
 For small magnitudes (small 
scales), one observes $\alpha \simeq 0.6 \pm 0.1$ 
that is usually interpreted as evidence of homogeneity ($D \approx 2$).
 At larger 
scales the value of $\alpha$ decreases to about $0.4$ (corresponding 
to $D \simeq 2$) 
and this deviation is considered as due to 
galaxy evolution and solid angle 
effects due to the expansion.
However, from the direct correlation analysis in 3-d space, we 
know that up to some rather large scale the galaxy 
distribution is certainly fractal. Therefore the usual interpretation of 
$\alpha \simeq 0.6 \pm 0.1$ 
at small scales in terms of a homogeneous distribution 
cannot be correct. In this respect the insert of 
Fig.\ref{fig3} allows us to clarify the 
situation. In fact, at relatively small scales,
 the density first raises and 
then falls when it reaches the correct scaling regime. This behaviour can 
roughly resemble a constant density, especially for an integrated 
quantity. A variety of tests for various catalogues with fractal 
correlations in 3-d, 
like the Perseus Pisces survey and others, show that the 
number counts give indeed an exponent $\alpha \simeq 0.6 \pm 0.1$
if the sample is dominated by small scales finite 
size fluctuations. On the other hand, if one computes the number counts 
by performing an average over all the different observers
in a redshift sample, the value $\alpha \approx 0.4$ 
is readily found, 
in agreement with the space correlation analysis. In the 
deep surveys ($ m \gtapprox 19$)
the small scale fluctuations are self-averaging so that one can 
recover the correct properties (i.e. $\alpha \approx 0.4$)
without performing any average. 

This implies a completely different 
interpretation of the number counts. At small scales the value 
$\alpha \approx 0.6$ is
 due to finite size effects and not to a real homogeneity, 
while at larger scales the value 0.4 corresponds to the correct 
correlation properties of the sample. This implies that galaxy evolution, 
modification of the Euclidean geometry and the K-corrections
 are not very relevant in the range of the present 
data. In addition the fact that the exponent 0.4 holds up to magnitudes 
$27 \div 28$
 seems to indicate that the fractal properties 
may extend up to $2 \div 
3000\hmp$. The amplitude of the galaxy number counts for $m \gtapprox 19$
 (see the solid line 
Fig.\ref{fig7}) is computed from the determination of the 
prefactor $B$ of the density $n(r) = Br^{D-3}$ at small scale
and from the knowledge of the galaxy luminosity function 
\cite{slgmp96}.
Quite a remarkable fact if one considers that the Hubble 
radius of the universe is supposed to be $4000\hmp$.
 Such a behaviour ($ \alpha \sim 0.4$, i.e. $D \sim 2$) 
can be found for galaxies in the 
different Photometric bands, 
as well as for other astrophysical objects \cite{slgmp96}.

We have now all the elements to reinterpret also 
the angular catalogues. These catalogues are qualitatively inferior 
to the 3-d ones because they only correspond to the angular projection 
and do not contain the third coordinate. However the fact that they 
contain more points with respect to the 3-d catalogues has led some 
authors to assign an excessive importance to these catalogues (once we 
have the 3-d ones).
In fact, even for a very large number of points 
($ N \rightarrow \infty$) the angular distribution is 
{\it intrinsically
 degenerate}, in the sense that it can originate from different
3-d distributions.
Actually the interpretation of the angular catalogues is quite 
delicate and ambiguous for a variety of reasons:
 
- Unlike orthogonal projections, angular projections mix different 
length scales and this gives an artificial randomization of the points. 
This can be illustrated as follows: consider a small solid angle 
$d \Omega$ of an angular catalog. This will contain the projection
of all the points in the corresponding cone of depth $R_d$. 
This cone is a three dimensional object, so the intersection 
with a fractal of dimension $D$ will also be a set of  dimension $D$. 
Therefore the number of points that project 
in $d \Omega$ is $N(d \Omega, R_d) \sim d \Omega R_d^D$. This 
implies that for a large enough value of $R_d$ any small solid angle 
$d\Omega$ will contain some points. This is why structures and voids
 are smeared by the angular projection.
This implies that the angular projection of a fractal structure can be 
really 
homogeneous at relatively large angles \cite{cp92} \cite{dp91} \cite{slmp97}. 
Clearly this is an artificial 
effect and from a smooth angular projection one cannot deduce 
whether the real distribution is also smooth. An example of this effect 
is given by Fig.2 in which the angular projection appears relatively 
smooth, while the real distribution is much more irregular. 
In fact the large structures and large voids have been identified 
with the redshift catalogues and could not be predicted from the 
angular data alone.

- An additional effect is the one due to the finite size effects discussed 
before in case of a single observational point that gives an additional 
artificial effect of homogenization. The so called rescaling of the 
angular correlations can be understood in detail by these effects and it 
can be shown that the same properties can be observed in a fractal 
distribution \cite{slgmp96}. 

Concerning the debate with M. Davis and the subsequent discussion 
with J. Peebles it should be noted that the angular projections 
correspond to 
a complex convolution with the magnitude limit and dilution effects in the 
case of IRAS. Contrary to the claim of M. Davis we have already 
analyzed the 
propreties 
of angular projections \cite{dp91} \cite{cp92} \cite{slmp97} and identified 
a number of problems, like those mentioned here, that were not known before
and that make these catalogues intrinsically unambiguous.
For these reasons we decided to focus on 3-d distributions 
that lead to the consistent and ambiguous results shown in Fig.\ref{fig3}.
Concerning the proposal of Davis and Peebles to generate a fractal 
distribution in 3-d that corresponds to the observed angular projections, this
requires the tuning of various other properties in addition to the 
fractal dimension. In particular lacunarity, morphology, magnitude limits and 
dilution effects all play a crucial role in the projection and 
should be tuned to those of real galaxies. In any case 3-d galaxy distributions 
are fractals and their properties do not depend on our ability 
to make this exercise.

\section{Luminosity and Space Distribution}

Up to now we have discussed galaxy correlations only in terms of the 
set of points corresponding to their position in space. Galaxies can be 
also characterized by their luminosity (related to their mass) and the 
full luminosity distribution is then a full distribution and not a simple 
set. It is natural then to consider the possible scale invariant 
properties of this distribution. This requires a generalization of the 
fractal dimension and 
the use of the concept of multifractality \cite{pie87}. In this 
language the fractal set discussed until now represents the support of 
the full luminosity distribution.

A multifractal analysis shows that also the full distribution is scale 
invariant \cite{cp92} \cite{pp96} 
and this leads to a new and important relation between the 
Schechter luminosity distribution and the space correlation properties. 
This allows us to understand various morphological features (like the 
fact that large elliptic galaxies are typically located in large clusters) in 
terms of multifractal exponents. This leads also to a new interpretation 
of what has been called the {\it luminosity segregation effect}
\cite{slp96}.

 The observation that the
 most luminous elliptical
galaxies lie in the core of rich galaxy clusters 
is, in our analysis,  a manifestation of the multifractal properties,
i.e. the self-similar distribution of matter including galaxy
masses (luminosities). However 
this observation together with the shift  of $r_0$ with sample depth
(interpreted as luminosity limit) has lead
various authors \cite{dav88} \cite{par94} \cite{ben96} 
to formulate the qualitative hypothesis 
that  the homogenization crossover (related to $r_0$)
should be different for galaxies of different luminosity.
The fact that large voids ($\gtapprox 50 \hmp$) 
are empty of galaxies of {\it any}
type is already a disproof of this hypothesis. 
In addition we have shown that the appropriate 
correlation  analysis
shows a power law behaviour at any observable scale.
This implies unavoidably that $r_0$ {\it must } scale with the sample
size because the system is self-similar. Most of the analysis of the
luminosity segregation effects usually do not address the 
fundamental question whether $r_0$ is a physical meaningful quantity,
that should be addressed with the conditional density.
Only once a crossover towards homogeneity would be observed in the 
conditional density then "luminosity segregation" questions could 
eventually be
 posed in the 
usual terms.

\section{Conclusions and Theoretical Implications}

In summary our main points are:
\begin{itemize}

\item
 The highly  irregular galaxy distributions with large structures and 
voids strongly point to a new statistical approach in which the 
existence of a well defined average density is not assumed a priori and 
the possibility of non analytical properties should be addressed 
specifically.

\item
 The new approach for the study
 of galaxy correlations in all the available catalogues 
shows that their properties are actually compatible with each other 
and they are statistically valid samples. The severe discrepancies 
between different catalogues that have led various authors to consider 
these catalogues as {\it not fair}, were due to the inappropriate methods of 
analysis.

\item
 The correct two point correlation analysis shows well defined fractal 
correlations up to the present observational limits, from 1 to 
$1000\hmp$ with fractal dimension $D \simeq 2$.
Of course the statistical quality and 
solidity of the results is stronger up to 
$100 \div 200 \hmp$ and 
weaker for larger 
scales due to the limited data. It is remarkable, 
however, that at these larger scales one observes exactly the continuation
of the correlation properties of the small and intermediate scales.

\item
 These new methods have been extended also to the analysis of the 
number counts and the angular catalogues which are shown to be fully 
compatible with the direct space correlation analysis. The new analysis of 
the number counts suggests that fractal correlations may extend also to 
scales larger that $1000\hmp.$

\item
The inclusion of the galaxy luminosity (mass) leads to a distribution 
which is shown to have well defined multifractal properties. This leads 
to a new, important relation between the luminosity function and that 
galaxy correlations in space.

\item
{\it New perspective on old arguments.}
On the light of these results we can now take 
a standard reference volume in the field (i.e. Peebles 1993 \cite{pee93})
and consider 
 the usual arguments invoked for homogeneity from a new point
of view. 
These arguments are: {\bf (a)} number counts: we have seen in Sec.5 that 
the small scale exponent of number counts is certainly not related to 
homogeneity but to small scale fluctuations. The real exponent of the number 
counts is instead the lower one (i.e. $\alpha \approx 0.4$) that indeed 
corresponds to the three dimensional 
(i.e fractal with $D \simeq 2$) 
correlation properties. {\bf (b) }
$\delta N/ N$ is
small for various observations. This point is exactly the same
as the fact $r_0$ is a spurious length. In the absence of 
a reference average one cannot talk about "large" or "small" 
amplitude of fluctuations. In addition, for any distribution, even
 a fractal one $\delta N/N$ is always small for sizes comparable
 to the total sample because the average is computed from the sample itself.
{\bf (c)} Angular correlations. We have seen in Sec.5 that angular correlations
are ambiguous in two respects: 
first the angular projection of a fractal is really
uniform at large angles due to projection effects, 
second the angular data are strongly affected by the finite size 
fluctuations that provide an additional artificial homogenization, as in the 
case of the number counts. The inclusion of these effects reconciles quite
 naturally the angular catalogs with the fractal properties in the three
dimensional ones. 
{\bf (d)} X-ray background. The argument that $\delta N/N$ becomes very small
for the X-ray background combines the two problems discusses before:
angular projections and reference average. This angular uniformity 
is analogues, for example, to the Lick angular sample, 
and certainly is
not a proof of real homogeneity. 

Finally one should note that all these arguments are {\it indirect}
and always require an interpretation based on some assumptions.
The most {\it direct} evidence for the properties of galaxy distribution 
arises from the correct correlation analysis of the 3-d volume limited samples
that has been the central point of our work.

\end{itemize}

{\bf Theoretical Implications}
\smallskip

From the theoretical point of view the fact that 
we have a situation characterized by {\it self-similar structures} 
 implies that we should not use concept
 like $\xi(r)$, $r_0$, $\delta N/N$ and certain properties of
the power spectrum, because they are not suitable to represent 
the real properties of the observed structures. 
In this respect also the N-body simulations
should be considered from a new perspective.
One cannot talk about "small" or "large" amplitudes 
for a self-similar structure because of the lack of a reference value like the 
average density.
The Physics should shift from {\it "amplitudes"} towards {\it "exponent" }
and the methods of modern statistical Physics should be adopted.
This requires the development of constructive interactions between two fields.
\bigskip

{\it Possible Crossover.}
We cannot exclude of course, that visible matter 
may really become homogenous at some large scale not 
yet observed. Even if this would 
happen the best way to identify the
eventual crossover is by 
using the methods we have described and  not the usual ones.
From a theoretical point of view  
the range of fractal fluctuations, extending at least over
three decades ($1 \div 1000 \hmp$), should  anyhow be
 addressed with the new theoretical concepts.
Then one should study the (eventual) crossover to homogeneity 
as an additional problem.
For the moment, however, no tendency to 
such a crossover is detectable from the experimental 
data and it may be reasonable to consider also more radical
 theoretical frameworks in which homogenization may 
simply not exist at any scale, at least for  
luminous matter.
\bigskip

{\it Dark Matter.} 
 All our discussion  refers to luminous matter. 
It would be nice if the new picture for the visible universe 
could reduce, to some extent, 
the importance of Dark matter in the theoretical framework.
At the moment however this is not clear. We have two possible 
situations: {\it (i)} if Dark matter is essentially 
associated to luminous matter, then the use of FRW metric is not
justified anymore.  This does not necessarily imply that there is 
no expansion 
or no Big Bang. 
It implies, however, that  these phenomena should be
described by more complex models. {\it (ii)}  If Dark matter
 is homogenous and luminous matter is fractal then, 
at large scale, Dark matter will dominate the gravity 
field and the FRW  metric is again valid. 
The visible matter however remains self-similar and non analytical 
and it still requires the new theoretical methods 
mentioned before. However, this perspective seems to be in contrast with the 
usual role of Dark Matter, that is to generate large potential wells 
for the initially smoother baryonic matter.
\bigskip

{\bf Predictions and Bets}
\smallskip

After the debate with Prof. M. Davis, we tried to assess 
the predictions of the two points of view and we also agreed to make a bet. 
According to the arguments of Prof. M. Davis, the length 
$r_0 \simeq 5 \hmp$ characterizes the physical properties of 
galaxy distributions. Therefore deeper samples like CfA2 and SLOAN
should simply reduce the error bar, which is 
now about considered 
to be $10 \%$. A possible variation of $r_0$ with absolute magnitude,
due to a luminosity bias (see Sec.6),  is 
considered plausible but it has 
never been quantified. This should be checked by
varying independently absolute magnitude and depth of the volume limited samples.
However, from this interpretation, the value of $r_0 = 5 \hmp$, 
corresponding to a volume limited of CfA1 with $M=-19.5$, should not change 
when considering in CfA2 and SLOAN volume limited samples with the 
same solid angle $\Omega$ and the same absolute magnitude limit ($M=-19.5$).

In our interpretation, instead,  $r_0$ is  spurious, and it scales 
linearly with the radius $R_s$ of the largest sphere fully contained 
in the volume limited samples. Therefore we predict for the
volume limited sample of CfA2 with $M=-19.5$ (with a solid angle of 
$\Omega=1.1 \; sr$ \cite{par94}) $r_0 \approx 7 \hmp$ (if, in the final version of the survey the solid angle will be $\Omega =1.8$, 
the value of $R_s$
will increase accordingly, and the value of $r_0$ will shift up 
to $\sim 9 \hmp$).
Note however that for the deepest 
volume limited CfA2 sample ($M \ltapprox -20$) we 
predict instead $r_{0} \approx 15 \div 20 \hmp$.
For the volume limited sample of the 
full SLOAN with $M=-19.5$
 ($\Omega = \pi$), 
our prediction is
that $r_0 \approx 65 \hmp$. It is clear that 
however, the first SLOAN slice will give smaller values 
because the solid angle will be small.

About the respective predictions for the full SLOAN ($\Omega = \pi \; sr$)
one of us 
(L.P.) made a bet with Prof. M. Davis \cite{dav96} (of a case of 
Italian or Californian wine). The predictions are 
$r_0 \simeq 5 \hmp$ (Davis) and $r_0 \gtapprox 50 \hmp$
(for the volume limited samples with $M<-19.5$), 
so the threshold for the bet 
was set (by the referee N. Turok)
to be the geometric average between the values: $r_0 =15 \hmp$.

\section*{Acknowledgments}
We thank for useful discussions, suggestions and collaborations
L.Amendola, A. Amici, D.J. Amit, Yu.V. Baryshev, 
R.Cappi, P. Coleman,  M.Davis, H. Di Nella,
 R. Durrer, A. Gabrielli, 
R.Giovanelli, B. Mandelbrot, G. Parisi, G. Paturel, 
P.J.E. Peebles, G. Salvini,  G. Setti, 
P. Teerikorpi, N. Turok, P. Vettolani and G. Zamorani.


\vfill\eject



\begin{thebibliography}{}   


\bibitem{cp92} Coleman, P.H. and Pietronero, L.,1992 Phys.Rep. 231,311

\bibitem{slmp97} Sylos Labini, F., Montuori, M.and 
 Pietronero,L.  1996, preprint

\bibitem{msla97} Montuori, M., Sylos Labini, F. and Amici, A. 
1996, preprint
 
\bibitem{pie87}Pietronero L., Physica A, 144, 257

\bibitem{pee93} Peebles, P.E.J., 1980 "The Large Scale Structure of 
The Universe" 
(Princeton Univ.Press.); Peebles, P.E.J.
1993 "Principles of physical
Cosmology" (Princeton Univ.Press.)


\bibitem{wei72} Weinberg, S. E. 1972
"Gravitation and Cosmology"
Wiley, New York.

\bibitem{wil82} Wilson K.G., 1974 Phys. Rep. 12, 75

\bibitem{amit86} Amit D., 1978
"Field theory, the Renormalization Group and Critical Phenomena
(Mc Graw-Hill, New York)


\bibitem{man83}Mandelbrot B., 1982 The Fractal Geometry of Nature,
Freeman, New York

   
\bibitem{epv95} Erzan A, Pietronero L., Vespignani A.
Rev. Mod. Phys. 1995, 67, 554
       
\bibitem{sl94}Sylos Labini, F., 1994, Ap.J.,  433, 464

\bibitem{del88} De Lapparent, V., Geller, M. J.,
Huchra, J. P. (1988) Ap.J.  332, 44
      
\bibitem{dp83} Davis, M., Peebles, P. J. E. 1983
Ap.J., 267,465

\bibitem{huc83} Huchra, J.,  Davis, M., Latham, D., Tonry, J.
 1983 Ap.J.S, 52, 89.


\bibitem{cps88}Coleman, P.H.  Pietronero, L.,\& Sanders,R.H.,1988, A\&A, 245,1


\bibitem{bs83} Bahcall N. A. \& Soneira R. M., 1983,
ApJ, 270, 20


\bibitem{bslmp94} Baryshev, Y., Sylos Labini, F.,
Montuori, M., Pietronero, L. Vistas in Astron. 1994, 38, 419
     


\bibitem{iras} Strauss M.A., {\em et al.}, 1992 Ap.J.S 83, 29; 
Fisher K. \etal 1995 Ap.J. Suppl. 100, 69
      
\bibitem{fis94} Fisher K., {\em et al} 1994 MNRAS 266, 507

\bibitem{dac88} Da Costa L.N., \etal 1991, ApJ. Suppl.,91, 935

\bibitem{lov96} Loveday J., 1996 MNRAS, 278, 1025

\bibitem{pp96} 
 Sylos Labini, F.,
  Montuori, M., Pietronero,L.  1996, Physica A, 230, 368; Di Nella H.,
Montuori M., Paturel G., Pietronero L.,
and Sylos Labini F., Astron.Astrophys.Lett. 1996, 308,  L33
                 
\bibitem{par94} Park, C., Vogeley, M.S., Geller, M., Huchra, J.
 1994
Ap.J., 431, 569;

\bibitem{cata} Haynes, M., Giovanelli, R., 1988 Large-scale motion 
in the Universe,
ed. Rubin, V.C., Coyne, G., Princeton University Press, Princeton;
 Paturel, G., Bottinelli, L., Gouguenheim, L., Fouque, P. 1988
A\&A, 189, 1;  Vettolani, G., {\em et al.} (1994)
 Proc. of Scloss Rindberg
workshop  Studying the Universe with Clusters
of Galaxies; Schectman \etal, 1996 Ap.J. 470, 172


\bibitem{bpslt96} Baryshev, Y., Pietronero, L, Sylos Labini, F.
\& Teerikorpi P., 1996 preprint

\bibitem{damsl96}   Amendola L, Di Nella H.,
Montuori M., 
and Sylos Labini F.,  1996 preprint
 
\bibitem{sla96} Sylos Labini, F. Amendola, L. 1996, Ap.J. Lett, 468, L1

\bibitem{slgmp96} Sylos Labini, F.,
Gabrielli, A., Montuori, M., Pietronero,L.  1996, Physica A, 266, 149

\bibitem{dp91} Dogterom M. and Pietronero L., 1991 Physica A, 171, 239

 
\bibitem{slp96}Sylos Labini, F., Pietronero,L. 1996. Ap.J. 469, 28

\bibitem{dav88} Davis M. \etal, 1988 ApJ Lett 333, L9

\bibitem{ben96} Benoist C. \etal 1996 ApJ in print

\bibitem{dav96} Davis M., in these Proceedings (astro-ph/9610149)






 
\end{thebibliography}
\end{document}